\documentclass[9pt,twocolumn,twoside]{osajnl}
\usepackage{soul} % Moriya added it for highlight
\usepackage{amsmath}% Moriya added it for multiline equations
\usepackage[justification=justified]{caption} % Moriya added it for alignment caption text to both sides
\usepackage{pdfpages}% Moriya added it for include supplementary

\journal{optica} % Choose journal (ao, aop, josaa, josab, ol, optica, pr)

\setboolean{shortarticle}{false}
\title{Acousto-optic Ptychography}

\author[1]{Moriya Rosenfeld}
\author[1]{Daniel Doktofsky}
\author[1]{Gil Weinberg}
\author[2]{Yunzhe Li}
\author[2]{Lei Tian}
\author[1,*]{Ori Katz}

\affil[1]{Applied Physics Department, Hebrew University of Jerusalem, 9190401 Jerusalem, Israel.}
\affil[2]{Department of Electrical and Computer Engineering, Boston University, Boston, MA 02215, USA}

\affil[*]{Corresponding author: orik@mail.huji.ac.il}

\begin{abstract}

Acousto-optic imaging (AOI) enables optical-contrast imaging deep inside scattering samples via localized ultrasound-modulation of scattered light. While AOI allows optical investigations at depths, its imaging resolution is inherently limited by the ultrasound wavelength, prohibiting microscopic investigations. 
Here, we propose a novel computational imaging approach that allows to achieve optical diffraction-limited imaging using a conventional AOI system. We achieve this by extracting diffraction-limited imaging information from 'memory-effect' speckle-correlations in the conventionally detected ultrasound-modulated scattered-light fields. 
Specifically, we identify that since speckle correlations allow to estimate the Fourier-magnitude of the field inside the ultrasound focus, scanning the ultrasound focus enables robust diffraction-limited reconstruction of extended objects using ptychography, i.e. we exploit the ultrasound focus as the scanned spatial-gate ‘probe’ required for ptychographic phase-retrieval. Moreover, we exploit the short speckle decorrelation-time in dynamic media, which is usually considered a hurdle for wavefront-shaping based approaches, for improved ptychographic reconstruction. 
We experimentally demonstrate non-invasive imaging of targets that extend well beyond the memory-effect range, with a 40-times resolution improvement over conventional AOI, surpassing the performance of state-of-the-art approaches.

\end{abstract}

\setboolean{displaycopyright}{false}

\begin{document}

\maketitle
%%%%%%%%%%%%%%%%%%%%%%% Introduction %%%%%%%%%%%%%%%%%%%%%%%%

\section{Introduction}
Optical microscopy through scattering media is a long-standing challenge with great implications for biomedicine. Since scattered light limits the penetration depth of diffraction-limited optical imaging techniques approximately to 1 millimeter, the goal of finding a better candidate for high-resolution imaging at depth is at the focus of many recent works \cite{ntziachristos2010going}. Modern techniques that are based on using only unscattered, ‘ballistic’ light, such as optical coherence tomography and two-photon microscopy, have proven very useful, but are inherently limited to shallow depths where a measurable amount of unscattered photons is present \cite{pawley2006handbook,webb1996confocal,huang1991optical,fercher2003optical,drexler2008optical,jang2020deep}.

The leading approaches for deep-tissue imaging, where no ballistic components are present, are based on the combination of light and ultrasound \cite{ntziachristos2010going}, such as acousto-optic tomography (AOT) \cite{elson2011ultrasound,wang2004ultrasound,resink2012state} and photoacoustic tomography (PAT) \cite{elson2011ultrasound,wang2012photoacoustic}). PAT relies on the generation of ultrasonic waves by absorption of light in a target structure under pulsed optical illumination. In PAT, images of absorbing structures are reconstructed by recording the propagated ultrasonic waves with detectors placed outside the sample. In contrast to PAT, AOT does not require optical absorption but is based on the acousto-optic (AO) effect: in AOT a focused ultrasound spot is used to locally modulate light at chosen positions inside the sample. The ultrasound spot is generated and scanned inside the sample by an external ultrasound transducer. The modulated, frequency-shifted, light is detected outside the sample using an interferometry-based approach \cite{elson2011ultrasound,resink2012state}. This enables the reconstruction of the light intensity traversing through the localized acoustic focus inside the sample. Light can also be focused back into the ultrasound focus via optical phase-conjugation of the tagged light in “time-reversed ultrasonically encoded” (TRUE) \cite{liu2015optical} optical focusing, or via iterative optimization, which can be used for fluorescence imaging \cite{wang2012deep,si2012fluorescence}. AOT and PAT combine the advantages of optical contrast with the near scatter-free propagation of ultrasound in soft tissues. However, they suffer from low spatial-resolution that is limited by the dimensions of the ultrasound focus, dictated by acoustic diffraction. This resolution is several orders of magnitude lower than the optical diffraction limit. For example, for ultrasound frequency of $50MHz$ the acoustic wavelength is $30 \mu m$, while the optical diffraction limit is $\frac{\lambda}{NA} $ where NA is the numerical aperture of the system and $\lambda$ is the optical wavelength, i.e. a 100-fold difference in resolution. This results in a very significant gap and a great challenge for cellular and sub-cellular imaging at depths.

In recent years, several novel approaches for overcoming the acoustic resolution limit of AOT based on wavefront-shaping have been put forward. These include iterative TRUE (iTRUE) \cite{si2012breaking,ruan2014iterative}, time reversal of variance-encoded (TROVE) optical focusing \cite{judkewitz2013speckle} and the measurement of the acousto-optic transmission matrix (AOTM) \cite{katz2019controlling}. 
%In TROVE the intensity fluctuations of ultrasonically tagged light for different random inputs are analyzed, and an optical wavefront that focuses on the location with increased fluctuations is shaped, allowing, in principle, optical speckle size focusing \cite{judkewitz2013speckle}. In iTRUE, multiple iterations of digital phase-conjugation are used to improve the focusing resolution \cite{si2012breaking, ruan2014iterative}. 
%%%
%%% 21.12.20 - MORIYA MADE IT COMMENT FOR SHORTER INTRODUCTION
%%%
Both iTRUE and TROVE rely on a digital optical phase-conjugation (DOPC) system \cite{cui2010implementation}, a complex apparatus, which conjugates a high-resolution SLM to a camera. In AOTM, an identical resolution increase as in TROVE is obtained without the use of a DOPC system, by measuring the transmission-matrix of the ultrasound modulated light and using its singular value decomposition (SVD)  for sub-acoustic optical focusing. A major drawback of this state-of-the-art 'super-resolution' AOT approaches is that they require performing a large number of measurements and wavefront-shaping operations in a time shorter than the sample speckle decorrelation time. In addition, in practice, these techniques do not allow a resolution increase of more than a factor of $\times3-\times6$ improvement from the acoustical diffraction-limit, when sub-micron optical speckle grains are considered \cite{katz2019controlling}. 
Recently, approaches that do not rely on wavefront shaping, and exploit the dynamic fluctuations to enable improved resolution \cite{doktofsky2020acousto} or fluorescent imaging \cite{ruan2020fluorescence} have been demonstrated, but these do not practically allow a resolution increase of more than a factor of 2-3. Closing the two orders of magnitude gap between the ultrasound resolution and the optical diffraction-limit is thus still an open challenge.

Diffraction-limited resolution imaging through highly scattering samples without relying on ballistic light is currently possible only by relying on the optical "memory effect" for speckle correlations \cite{feng1988correlations,bertolotti2012non,katz2014non}. These techniques retrieve the scene behind a scattering layer by analyzing the correlations within the speckle patterns. Unfortunately, the memory-effect has a very narrow angular range, which limits these techniques to isolated objects that are contained within the memory-effect field-of-view (FoV). For example, at a depth of $1mm$ the memory-effect range is of the order of tens of microns \cite{judkewitz2015translation,osnabrugge2017generalized} making it inapplicable for imaging extended objects.

Here, we present acousto-optic ptychographic imaging (AOPI), an approach that allows optical diffraction-limited imaging over a wide FoV that is not limited by the memory-effect range, by combining acousto-optic imaging (AOI) with speckle-correlation imaging. 
Specifically, we utilize the ultrasound focus as a controlled probe that is scanned across the wide imaging FoV, and use speckle-correlations to retrieve optical diffraction-limited information from within the ultrasound focus. Importantly, we develop a reliable and robust computational reconstruction framework that is based on ptychography \cite{maiden2009improved, maiden2017further,pham2019semi}, which exploits the intentional partial overlap between the ultrasound foci.
We demonstrate in a proof of principle experiments a $>\times 40$ increase in resolution over the ultrasound diffraction-limit, providing a resolution of $3.65\mu m$ using a modulating ultrasound frequency of $25MHz$.
%%%%%%%%%%%%%%%%%%%%%%% Principle %%%%%%%%%%%%%%%%%%%%%%%%%%%%%
\section{Methods}
\subsection{Principle}
\begin{figure*}[t!]
\centering\includegraphics[width=0.9\linewidth]{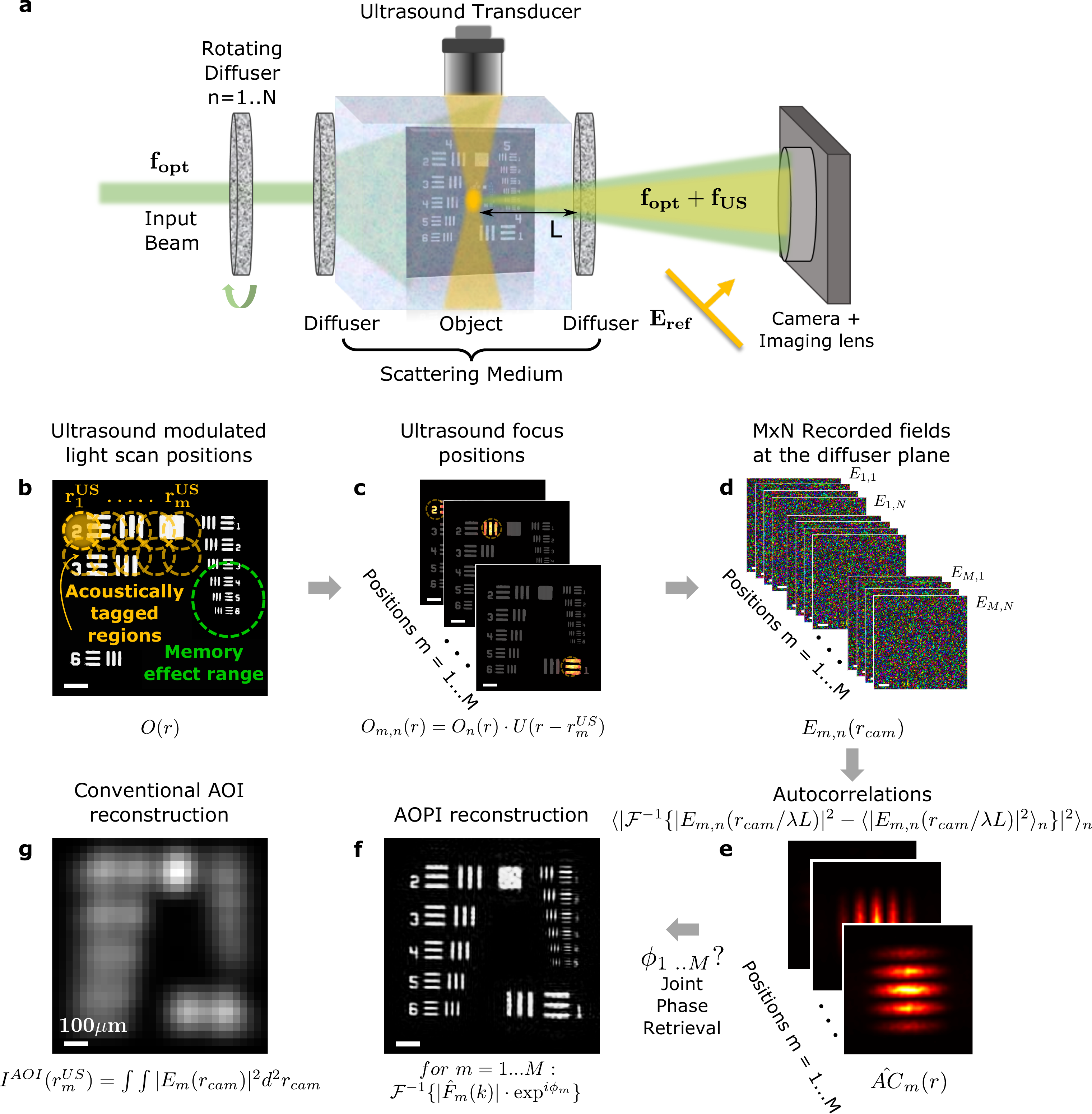}
\caption{Acousto-optic ptychographic imaging (AOPI) principle and numerical results.  a. Schematic of the experimental setup: An AOI setup is equipped with a rotating diffuser for producing controlled speckle realizations. An object hidden inside a scattering sample is imaged by scanning a focused ultrasound beam (in yellow) over the object, and the acousto-optic modulated (frequency-shifted) light is holographically detected using a high-resolution camera. b. The ultrasound beam scan the target (dashed yellow circle). The US beam is smaller than the "memory effect" range (dashed green circle), allowing the use of speckle correlation imaging for each scan as part of our AOPI method. 
c. For $m = 1...M$ scan positions, the ultrasound beam modulate the light at the target plane.  
d.The modulated light propagate from the target plane through a diffuser and reach the camera. For each scan position, N different speckle realizations fields are recorded at the camera, due to different speckle illuminations that are obtained using the rotating diffuser.
e. For each scan position, the autocorrelation of the ultrasound modulated light is estimated via correlography, using the N recorded fields. 
f. Numerical result for AOPI reconstruction. The M autocorrelations for all scan positions are entered into ptychography-based phase retrieval algorithm and a full reconstruction of the target is obtained. 
g. Conventional AOI reconstruction, obtained by plotting the total modulated power at each ultrasound focus position. 
Scale bar $100\mu m$.}
\label{Figure1}
\end{figure*}

The principle of our approach is presented in Figure \ref{Figure1} along with a numerical example. Our approach is based on a conventional pulsed AOI setup, employing a camera-based holographic detection of the ultrasound modulated light \cite{elson2011ultrasound,katz2019controlling}. In this setup  (Fig. \ref{Figure1}(a)) the sample is illuminated by a pulsed quasi-monochromatic light beam at a frequency $f_{opt}$. The diffused light is ultrasonically tagged at a chosen position inside the sample by a focused ultrasound pulse at a central frequency $f_{US}$. The acousto-optic modulated (ultrasound-tagged) light field at frequency $f_{AO}=f_{opt}+f_{US}$ is measured by a camera placed outside the sample using a pulsed reference beam that is synchronized with the ultrasound pulses, via off-axis phase-shifting interferometry \cite{doktofsky2020acousto,gross2005heterodyne} (Supplementary section 1).

In conventional AOI, the ultrasound focus is scanned along the target object (Fig. \ref{Figure1}(b,c)), and the AOI image, $I^{AOI}(\bold{r})$, is formed by summing the total power of the detected ultrasound-modulated light at each ultrasound focus position $r^{US}_{m}$: $I^{AOI}(r^{US}_m)=\iint{|E_m(\bold{r_{cam}})|^2}d^2r_{cam}$, where $E_m(\bold{r_{cam}})$ is the ultrasound-modulated speckle field that is measured by the camera (Fig. \ref{Figure1}(d)). Since the conventional AOI image (Fig. \ref{Figure1}(g)) is a convolution between the target object and the ultrasound-focus pressure distribution \cite{doktofsky2020acousto}, its resolution is limited by the acoustic diffraction limit.

Our approach relies on the same data acquisition scheme as in conventional AOI (Fig. \ref{Figure1}(b-d)). However, instead of integrating the total power of the camera-detected ultrasound-modulated field at each ultrasound position, we use the spatial information in the detected field, $E_m(\bold{r_{cam}})$, to reconstruct the diffraction-limited target features \textit{inside} the ultrasound focus, via speckle-correlation computational imaging approach  %(Fig.\ref{Figure1}f)
\cite{bertolotti2012non,katz2014non,edrei2016optical}. Specifically, we estimate the autocorrelations of the hidden target inside each ultrasound focus position  (Fig. \ref{Figure1}(e)), and then use a ptychography-based algorithm \cite{maiden2009improved,maiden2017further} to jointly reconstruct the entire target from all estimated autocorrelations (Fig. \ref{Figure1}(f)). 
Thus, our approach exploits the richness in the information of the detected ultrasound-modulated speckle fields, which contain a number of speckle grains limited only by the camera pixel count. 

Beyond improving the resolution of AOI by several orders of magnitude, from the ultrasound diffraction-limit to the optical diffraction limit, our approach allows to tackle a fundamental and generally very difficult to fulfill a requirement for speckle-correlation imaging: that the entire imaged object area must be contained within the memory-effect correlation-range \cite{freund1990looking,katz2014non,bertolotti2012non}, i.e. that all object points produce correlated speckle patterns \cite{katz2014non}. This requirement usually limits speckle-correlation imaging to small and unnaturally-isolated objects. 
Recently, ptychography-based approaches were utilized to overcome the memory-effect FoV  \cite{gardner2019ptychographic,li2019imaging}. However, the implementations of all FoV-extending approaches to date required direct access to the target in order to limit the illuminated area, a requirement that is impossible to fulfill in noninvasive imaging applications.
Our approach overcomes this critical obstacle by relying on noninvasive ultrasound tagging to limit the detected light to originate only from a small controlled volume that is determined by the ultrasound focus. The only requirement for speckle-correlation imaging is that the ultrasound focus (Fig. \ref{Figure1}(b), dashed yellow circle) would be smaller than the memory-effect range (Fig. \ref{Figure1}(b), dashed green circle). Thus allowing to image through scattering layers objects that extend well beyond the memory-effect FoV, without a limit on their total dimensions.  

%%%%%%%%%%%%%%%%%%%%%%% Figure1 %%%%%%%%%%%%%%%%%%%%%%%%%%%%%

%%%%%%%%%%%%%%%%%%%%%%% end Figure1 %%%%%%%%%%%%%%%%%%%%%%%%%%%%%ֿ
Mathematically our approach can be described as follows: Consider a target object located inside a scattering sample (Fig. \ref{Figure1}(a)). As a simple model, we model the object by a thin 2D amplitude and a phase mask, whose complex field transmission is given by $O(\bold{r})$. The goal of our work is to reconstruct the object 2D transmission $|O(\bold{r})|^2$ by noninvasive measurements of the scattered light distributions outside the sample.  A monochromatic spatially-coherent laser beam illuminates the object through the scattering sample (Fig. \ref{Figure1}(a)). The light propagates through the scattering sample, results in a speckle illumination pattern at the object plane. Considering a dynamic scattering sample, such as  biological tissue, the illuminating speckle pattern on the object is time-varying. We denote the speckle pattern field illuminating the object at a time $t_n$ by $S_n(\bold{r})$.
The field distribution of the light that traverses the object at time $t_n$ is thus given by: $O_n(\bold{r})=O(\bold{r})S_n$.  This light pattern is ultrasound modulated by an ultrasound focus whose central position,  $r^{US}_{m}$ , is scanned over $m=1..M$ positions inside the sample. We denote the ultrasound focus pressure distribution at the m-th position by $U(\bold{r}-r^{US}_m)$. The shift-invariance of the ultrasound focus is assumed here for simplicity of the derivation, and is not a necessary requirement \cite{peterson2016probe}. 
The ultrasound modulated light field at the m-th ultrasound focus position is given by the product of $O_n(\bold{r})$ and $U(\bold{r}-r^{US}_m)$: $O_{m,n}(\bold{r})=O_n (\bold{r})U(\bold{r}-r^{US}_m )$ . The ultrasound modulated light field $O_{m,n}(\bold{r})$ propagates to the camera through the scattering sample, producing a random speckle field at the camera plane: $E_{m,n}(\bold{r_{cam}})$. 
When the ultrasound focus dimensions are smaller than the memory-effect range:  $D_{US}<\Delta r_{mem}\approx L\Delta \theta_{mem}$, where $L$ is the depth of the object inside the scattering sample from the camera side (Fig. \ref{Figure1}(a)) and $\Delta \theta_{mem}$ is the angular range of the memory effect, the scattering sample can be considered as a thin random phase-mask with a phase distribution $\phi_{sample}(\bold{r_{cam}})$. The scattered light field measured by a camera that images the scattering sample facet is thus given by: 
\begin{equation}
E_{m,n}(\bold{r_{cam}}) =\mathcal{P}_{L}[O_{m,n}(\bold{r})]exp(i\phi_{sample}(\bold{r_{cam}}))
\label{propogation}
\end{equation}
 where $\mathcal{P}_{L}$ is the propagation operator for propagating the field from the object plane to the scattering sample facet. 
The complex field autocorrelation of the target, illuminated by the n-th speckle pattern and multiplied by the ultrasound field distribution  $ O_{m,n}(\bold{r}) \star O_{m,n}(\bold{r})$  can be calculated from a single camera frame \cite{idell1987image}. However, multiple ($n=1..N$) camera frames, captured under different speckle illuminations, can be used to calculate the target intensity autocorrelation: $ {AC}_m(\bold{r})=|O_m(\bold{r})|^2\star |O_m(\bold{r})|^2$, which is free from speckle artefacts via correlography \cite{idell1987image}, in the exact same manner as demonstrated without ultrasound modulation by Edrei et al \cite{edrei2016optical}. 
In correlography \cite{idell1987image,edrei2016optical,li2019imaging} the estimate for the object un-speckled intensity autocorrelation at the m-th ultrasound focus position, $\hat{AC}_m(\bold{r})$, is calculated by averaging the Fourier transforms of the captured speckle frames intensity distribution, after subtracting their mean value \cite{idell1987image,  li2019imaging}:

\begin{equation}
\begin{aligned}
  \hat{AC}_m(\bold{r})=\langle|\mathcal{F}^{-1}\{|E_{m,n}(\bold{r_{cam}}/\lambda L)|^2-\langle |E_{m,n}(\bold{r_{cam}}/\lambda L)|^2 \rangle_{n}\} |^2 \rangle_n \\
  = \underbrace{[|O_m(\bold{r})|^2\star |O_m(\bold{r})|^2] }_\text{$AC_m(\bold{r})$} * \underbrace{\langle|S_n(\bold{r})\star S_n(\bold{r})|^2\rangle_n}_{\substack{\textrm{speckle grain}\\  \textrm{autocorrelation}}}
\end{aligned}
\label{correlography}
\end{equation}

The calculated autocorrelation, $\hat{AC}_m(\bold{r})$, is the autocorrelation of the object convolved with the diffraction-limited point-spread-function of the imaging aperture on the facet plane. A required condition for estimating the object autocorrelation from the Fourier-transform of the captured speckle patterns (Eq.\ref{correlography}) is that the object distance from the measurement plane, $L$, is larger than $2Dr_c/\lambda$, where D is the object dimensions, i.e. the ultrasound focus diameter, and $r_c$ is the illumination speckle grain size \cite{edrei2016optical}. In deep tissue imaging, $r_c\approx \lambda/2$, and the condition becomes: $L>D$, i.e. the imaging depth should be larger than the dimensions of the ultrasound focus, a naturally fulfilled condition.

The use of a multiple illumination speckle realizations, $N$, is advantageous both for estimating the un-speckled intensity transmission of the object, $|O_m(\bold{r})|^2$ , and for improved ensemble averaging of the estimation \cite{katz2014non,edrei2016optical} (see Supplementary section 5). For a dynamic sample, the speckle illuminations naturally vary in time, and in the case of a static sample, the speckle realizations can be easily obtained by e.g. a rotating diffuser. 
According to the Wiener-Khinchin theorem, the Fourier transform of the object estimated autocorrelation, $\hat{AC}_m(\bold{r})$, is the Fourier magnitude of the object intensity transmission:  $|\hat{F}_m(\bold{k})|= \sqrt{\mathcal{F}\{\hat{AC}_m(\bold{r})\}}$
The object itself can thus be reconstructed from $|\hat{F}_m(\bold{k})|$ via phase retrieval \cite{fienup1978reconstruction}. 
\begin{equation}
for\ m=1...M: |O_m(\bold{r})|^2=\mathcal{F}^{-1} \{|\hat{F}_m(\bold{k})|\exp(i\phi_m)\} 
 \label{eq1.9}
\end{equation}
If a partial overlap between the scanned ultrasound foci exists, the reconstruction problem can be reliably solved using ptychography, an advanced joint phase-retrieval technique \cite{rodenburg2004phase,maiden2009improved,maiden2017further,pham2019semi}, that was recently shown to be extremely successful in stable, high-fidelity, robust reconstruction of complex objects, which is not possible by separately solving the $M$ phase-retrieval problem. 
A numerical example for using our approach to image an extended object beyond the memory-effect FoV, with diffraction-limited resolution is shown in Fig \ref{Figure1}.f, side-by-side with the conventional AOI image of the same object using the same measurements (Fig. \ref{Figure1}(g)). A resolution increase of  $>\times 25$ over conventional AOI is apparent in the high-fidelity reconstruction (Fig. \ref{Figure1}(f)). A detailed explanation of the data processing and the implemented ptychographic reconstruction algorithm can be found in Supplementary section  4,7. 

%%%%%%%%%%%%%%%%%%%%%%% Results %%%%%%%%%%%%%%%%%%%%%%%%%%%%%

\section{Results}
\subsection{Experimental set up}
To demonstrate our approach we built a proof-of-principle setup schematically shown in Fig. \ref{Figure1}(a). It is a conventional AOI setup with camera-based holographic detection based on phase-shifting off-axis holography (Supplementary section 1), with the addition of a controlled rotating diffuser before the sample (two $1^o$ light shaping diffusers, Newport), used for generating dynamic random speckle illumination. The illumination is provided by a pulsed long-coherence laser at a wavelength of 532nm (Standa). An ultrasound transducer with a central frequency of $f_{US}=25MHz$, and ultrasound focus dimensions of $D_X=149\mu m$, $D_Y=140\mu m$ full-width at half max (FWHM) in the horizontal (transverse) and vertical (axial ultrasound) directions, correspondingly,  is used for acousto-optic modulation. The ultrasound focus position was scanned laterally by a motorized stage, and axially by electronically varying the time delay between the laser and ultrasound pulses. The full setup description is given in Supplementary Section 1.
As controlled scattering samples and imaged targets for our proof-of-principle experiments we used a sample comprised of a target placed in water between two scattering layers composed of several $5^o$ scattering diffusers that have no ballistic component (see Supplementary Section 6). An sCMOS camera (Andor Zyla 4.2 plus) was used to holographically record the ultrasound-modulated scattered light fields, using a frequency-shifted beam. To minimize distortions in the recorded fields, no optical elements were present between the camera and the diffuser. The field at the diffuser plane, $E_{m,n}(\bold{r_{cam}}/\lambda L)$, was calculated from the camera recorded field by digital propagation (Supplementary section 1).

%%%%%%%%%%%%%%%%%%%%%%% Figure2 %%%%%%%%%%%%%%%%%%%%%%%%%%%%%

\begin{figure*}[h!]
\centering\includegraphics[width=0.85\linewidth]{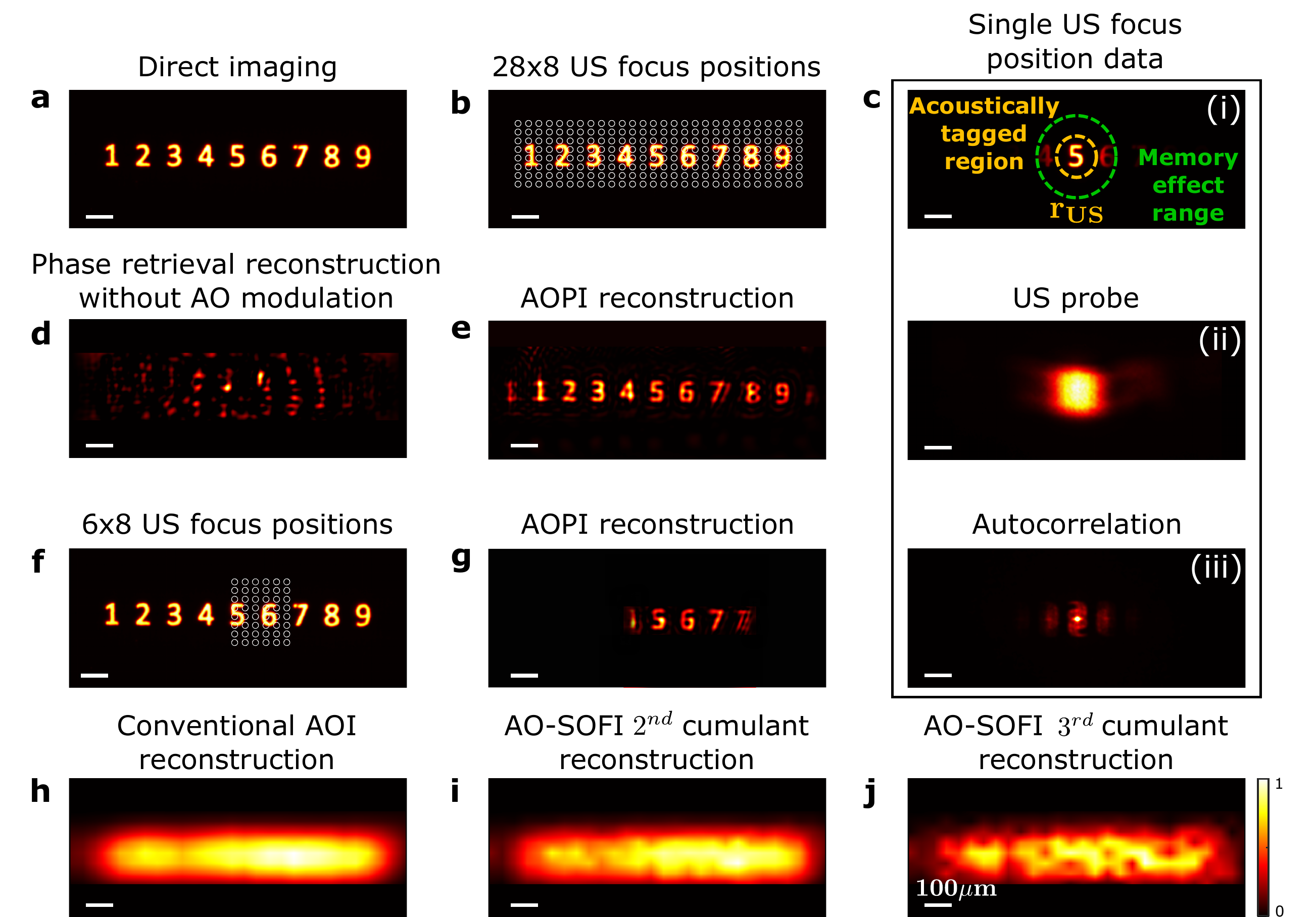}
\caption{Experimental AOPI result of an extended object beyond the memory-effect. a. Direct imaging of the target object. b. Acoustic focus centers of the  $28\times8$ scan positions (white circles). c. Single acoustic focus position data. (i) While the "memory effect" region (dashed green circle) is smaller than the object dimension, preventing the possibility to use speckle correlation method for the entire object, it is bigger than the acoustically tagged region (dashed green circle).  (ii).The acoustic probe. Probe dimensions: horizontal FWHM $149\mu m$, vertical FWHM $140\mu m$. (iii). Single estimated autocorrelation of c(i). d. Speckle correlation reconstruction using classic phase retrieval algorithm (same algorithm as in the ptychography engine). e. AOPI reconstruction using rPIE algorithm. f-g. Same as b,e respectively, for only $6\times8$ probe positions, where the environment of the scanned object is not isolated. h. Conventional AOI reconstruction. The acoustic probe limits the imaging resolution to the probe dimensions. i. AO-SOFI $2^{nd}$ cumulant reconstruction. j. AO-SOFI $3^{rd}$ cumulant reconstruction. h-j reconstruction are based on the same  $28\times8$ scan positions measurements. Scale bar $100\mu m$.}
\label{Figure2}
\end{figure*}

%%%%%%%%%%%%%%%%%%%%%%% end Figure2 %%%%%%%%%%%%%%%%%%%%%%%%%%%%%

\subsection{Imaging an extended object beyond the memory effect }

As a first demonstration, we imaged a transmittive target composed of nine digits (Fig. \ref{Figure2}(a)) that extends over $3.5$ times beyond the memory-effect of the scattering sample, which is $\Delta r_{mem} \sim 280 \mu m$ (Fig. \ref{Figure2}(c(i)), dashed green circle, Supplementary Fig. S3).

For 2D imaging, the ultrasound focus (Fig. \ref{Figure2}(c(ii))) was scanned over the target with a step size of $\Delta X=44.7\mu m$ ,$\Delta Y=37.3\mu m$, in the horizontal and vertical directions, respectively. These steps (along with the ultrasound spot size) define a probe overlap of  $\sim 88\%$ between neighboring  positions (Fig. \ref{Figure2}(b)). A study of the effect of the probe overlap on the reconstruction fidelity is presented in Supplementary section 8.
For each ultrasound focus position, $r^{US}_m$ ($m=1..224$) we recorded $N=150$ different ultrasound-modulated light fields, $E_{m,n}(\bold{r_{cam}})$, each with a different (unknown) speckle realization, $S_n(\bold{r})$. 
The target was reconstructed from the $M=224$ autocorrelations using rPIE ptychographic algorithm \cite{maiden2017further} (Supplementary section 7). Fig. \ref{Figure2}(c(iii)) presents an example for one of the autocorrelations used as input. 
The AOPI reconstructed image (Fig. \ref{Figure2}(e)) provide an image with a resolution well beyond that of a conventional AOI reconstruction (Fig. \ref{Figure2}(h)), and also well beyond the improved resolution of recent super-resolution AOI techniques such as AO-SOFI \cite{doktofsky2020acousto} (Fig. \ref{Figure2}(i-j)). 
Importantly, since the target extends beyond the memory-effect range, as is the case in many practical imaging scenarios, conventional speckle-correlation imaging without AO modulation \cite{edrei2016optical,katz2014non} fail to reconstruct the object (Fig. \ref{Figure2}(d)), as expected.  Interestingly, when taken only part of the measured positions from the center of the scanned area (only $6\times8$ positions from the center of the object area, instead of all $28\times8$  positions, Fig. \ref{Figure2}(f)), and reconstruct the probed object - a good reconstruction is obtained using AOPI method (Fig. \ref{Figure2}(g)), proving that the acoustic probe well functions as an isolation aperture.

%%%%%%%%%%% Super resolution AOI of USAF-1951 portion %%%%%%%%%

\subsection{Imaging resolution verification}
%%%%%%%%%%%%%%%%%%%%%%% Figure3 %%%%%%%%%%%%%%%%%%%%%%%%%%%%%

\begin{figure*}[h]
\centering\includegraphics[width=0.85\linewidth]{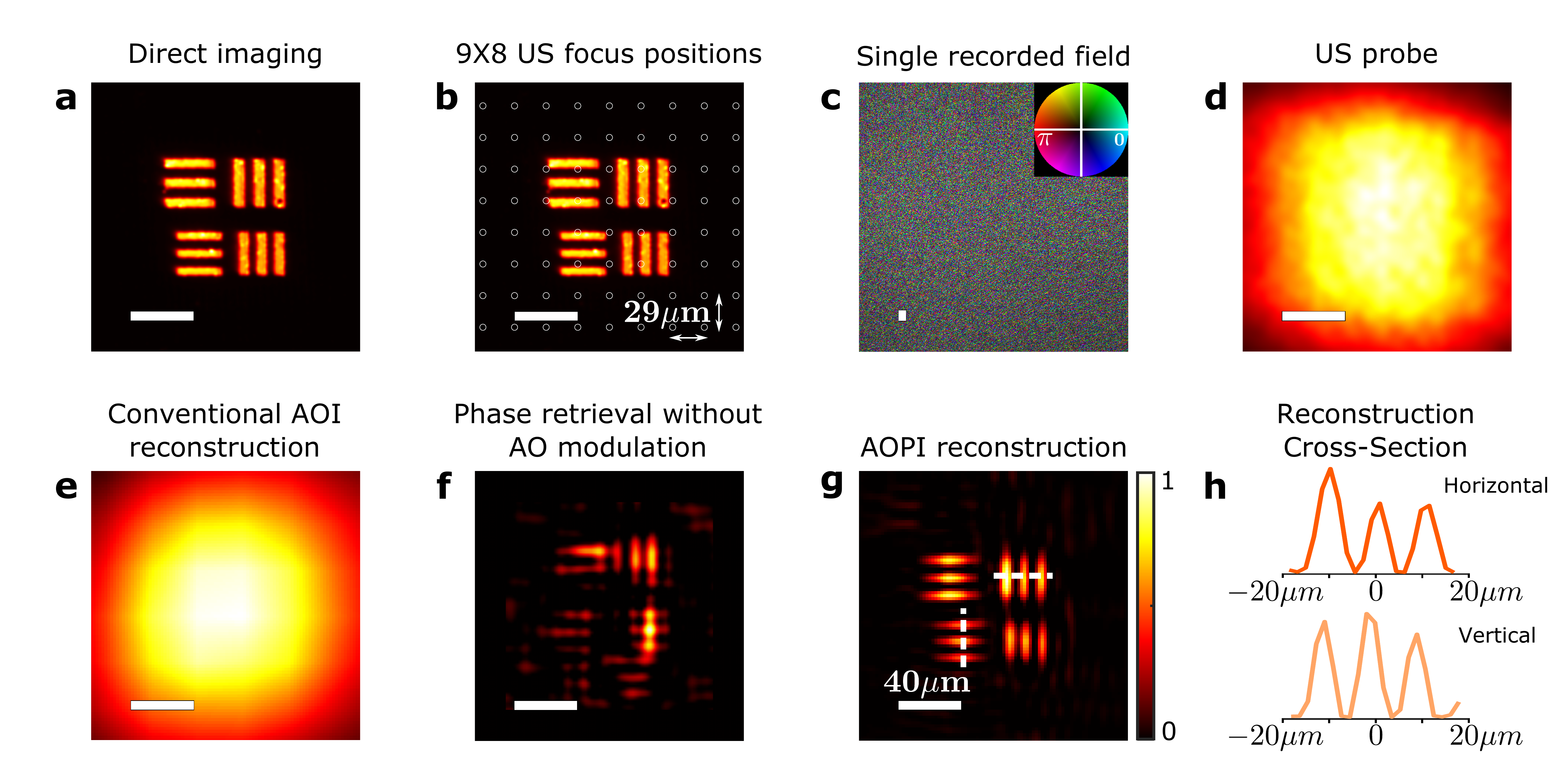}
\caption{Experimental AOPI result of a high-resolution target. a. Direct imaging of the target object. The object is smaller than the acoustic probe and the "memory effect" range.  b. Probe centers of the $9\times8$ scan positions (white circles). c. Single ultrasound-modulated light recorded field at the camera plane. d. The acoustic probe. Probe dimensions: horizontal FWHM $149\mu m$, vertical FWHM $140\mu m$. e. Conventional AOI reconstruction. The acoustic probe limits the imaging resolution to the probe dimensions. f. Speckle correlation reconstruction using classic phase retrieval algorithm (same algorithm as in the ptychography engine). g. AOPI reconstruction using rPIE algorithm. h. Cross-section of the AOPI reconstruction from (g), resolve features $\sim \times30$  smaller than the acoustic focus with  $\sim \times40$  increase in resolution compared to classical AOI methods.  Horizontal cross-section lines width is $6.2\mu m$ (orange) and vertical cross-section lines width is $5.52\mu m$ (bright orange). Scale bar $40 \mu m$.}
\label{Figure3}
\end{figure*}

%%%%%%%%%%%%%%%%%%%%%%% end Figure3 %%%%%%%%%%%%%%%%%%%%%%%%%%%%%
To demonstrate the resolution increase of AOPI  we performed an additional experiment where the target of Fig. \ref{Figure2} was replaced by elements 3-4 of group 6 of a negative USAF-1951 resolution test chart (Fig. \ref{Figure3}(a)). For 2D imaging, the ultrasound focus ( Fig. \ref{Figure3}(d)) was scanned over the target with a step size of $\Delta X=\Delta Y=29\mu m$. These steps (along with the ultrasound spot size) define a probe overlap of  $\sim 93\%$ between neighbouring positions (Fig. \ref{Figure3}(b)). For each ultrasound focus position, $r^{US}_m$, $m=1..72$ we recorded $N=150$ different ultrasound-modulated light fields, $E_{m,n}(\bold{r_{cam}})$  (Fig \ref{Figure3}.c), each with a different (unknown) speckle realization, $S_n(\bold{r})$. The reconstruction of the target from the $M=72$ autocorrelations using rPIE ptychography algorithm \cite{maiden2017further} is presented in Fig. \ref{Figure3}(g-h). A study of the effect of probe overlap on the reconstruction is presented in Supplementary section 8. The AOPI reconstructed image resolves resolution target features of size (separation) of $5.52\mu m$ (Fig. \ref{Figure3}(g-h)), which is   $\sim \times30$  smaller than the acoustic focus FWHM ($\sim 145\mu m$). 
The cross-sections of the reconstructed image (Fig. \ref{Figure3}(h)) allow to estimate the imaging resolution by fitting the result to a convolution of the known sample structure with a Gaussian PSF. This results in a resolution of $ 3.65\mu m$ (FWHM), a  40-fold increase in resolution compared to the acoustic resolution of conventional AOI (Fig. \ref{Figure3}(e)). 
Interestingly, although the target dimensions in this experiment are contained in the memory-effect range, conventional speckle-correlation imaging that is based on phase-retrieval without AO modulation (Fig. \ref{Figure3}(f)), results in a considerably lower reconstruction fidelity than the AOPI reconstruction (Fig. \ref{Figure3}(g)). This improvement is attributed to the larger input data set and improved algorithmic stability of ptychographic reconstruction compared to simple phase-retrieval \cite{rodenburg2004phase,maiden2009improved}.
%%%%%%%%%%%%%%%%%%%%%%% Discussion %%%%%%%%%%%%%%%%%%%%%%%%%%%%%

\section{Discussion and conclusion}
We proposed and demonstrated an approach for diffraction-limited wide FoV optical imaging in ultrasound with speckle-correlation computational imaging \cite{edrei2016optical}. In contrast to previous approaches for super-resolved acousto-optic imaging \cite{judkewitz2013speckle,katz2019controlling,ruan2014iterative, si2012breaking}, the resolution of our approach is optically diffraction-limited, independent of the ultrasound probe dimensions, the ratio between the speckle grain size and the ultrasound probe dimensions, or the number of realizations\cite{katz2019controlling}. 
This allowed us to demonstrate an $\times 40$ improvement in resolution over the acoustic diffraction-limit, an order of magnitude larger gain in resolution compared to the state of the art approaches, such as iTRUE \cite{ruan2014iterative,si2012breaking}, TROVE \cite{judkewitz2013speckle}, and AOTM \cite{katz2019controlling}. In addition, TROVE and AOTM allow the resolution to increase only when unrealistically large speckle grains are considered \cite{katz2019controlling}. 

Another important advantage of our approach is that unlike transmission-matrix and wavefront-shaping based approaches \cite{judkewitz2013speckle,katz2019controlling} it does not require unrealistically long speckle decorrelation times. Similar to recent approaches that utilize random fluctuations \cite{doktofsky2020acousto,li2019imaging}, our approach benefits from the natural speckle decorrelation to generate independent realizations of coherent illumination, improving the estimation of the object autocorrelation \cite{edrei2016optical}.

While our approach relies on the memory-effect to retrieve the diffraction-limited image, its FoV is not limited by the memory-effect range, as is the case in all other noninvasive memory-effect based techniques \cite{bertolotti2012non,katz2014non}. The FoV is dictated by the scanning range of the ultrasound focus, which is practically limited only by the allowed acquisition time. Such an extension of the FoV in speckle-correlation based imaging has only been obtained before by invasive access to the target object \cite{li2019image,gardner2019ptychographic,zhou2020retrieval,divitt2020imaging}.  The adaptation of a ptychographic image reconstruction significantly improves the reconstruction fidelity and stability compared to phase-retrieval reconstruction (Fig. \ref{Figure3}(f-g)) \cite{katz2014non,bertolotti2012non}.

Our super-resolution AOPI approach does not rely on wavefront-shaping \cite{judkewitz2013speckle,si2012breaking,ruan2014iterative,katz2019controlling} or nonlinear effects\cite{selb2002nonlinear}, and it can be applied to any AOI system employing camera-based coherent detection.

The main limitation of our approach is the requirement for a memory-effect range which is of the order of the ultrasound probe dimensions, i.e. that the ultrasound-tagged fields have a non-negligible correlations. This condition can be satisfied by relying on a small ultrasound focus, achieved by the use of high-frequency ultrasound, and by the use of long laser wavelength, which increases the memory-effect range \cite{freund1988memory,freund1990looking}.
Importantly, while at very large imaging depths, deep within the diffusive light propagation regime, the memory effect angular range is $\theta _{mem}\approx \frac{\lambda}{L}$ \cite{freund1988memory,freund1990looking}, at millimeter-scale depths, which are of the order of the transport mean free path (TMFP), the memory-effect range has been shown to be orders of magnitude larger \cite{schott2015characterization}. In addition, the requirement for a sufficiently large memory-effect can be alleviated by relying on translation-correlations \cite{judkewitz2015translation} or the generalized memory-effect \cite{osnabrugge2017generalized}.
Applying the above improvements (a higher ultrasound frequency, a longer optical wavelength, generalized speckle correlations) are the next steps for bringing our proof of principle demonstrations to practical biomedical imaging applications.

Another necessary technical improvement for biomedical application of our approach is in the demonstrated acquisition speed. Similar to all super-resolution techniques that do not rely on object priors, our approach requires a large number of measurements for reconstructing a single image. The required number of frames is the product of: (the number of probe positions) $\times$ (the number of realization per probe position)  $\times$ (number of phase-shifting frames). In our proof of principle demonstration system, which was not optimized for acquisition speed, we used 150 realizations per probe position and 16 phase-shifting frames. A diffuser mounted on a slowly rotating motor was used to change the realizations, and a conventional sCMOS camera was used to capture the images, which resulted in an acquisition time of $\sim 1.6$  seconds per realization. This time can be reduced by orders of magnitude using single-shot, off-axis, fast camera-based detection \cite{liu2017focusing,katz2019controlling, liu2016lock,ruan2020fluorescence}, a fast MEMS-based dynamic wavefront randomizer \cite{shevlin2018phase},  and 2D electronic scanning US array instead of the mechanical scan of the single-element ultrasound transducer \cite{laudereau2016ultrafast}. Assuming the acquisition speed is limited by the camera frame-rate of $\sim 7,000$ frames-per-second \cite{katz2019controlling}, the acquisition of  150 realizations for 72 probe positions (as in Fig. \ref{Figure3}), will be of the order of  $\sim 1.5sec$, excluding off-line data processing. This is expected to be adequate for imaging biological structures since the requirement on the speckle decorrelation time is that of a single frame, i.e. $\sim 0.1msec$. Moreover, the number of required realizations is expected to be significantly reduced by using advanced correlation-based reconstruction schemes such as those provided by deep neural networks (DNN). These have been recently shown to be able to significantly improve the estimation of the intensity autocorrelation, from only a few coherent realizations \cite{metzler2020deep}.

The combination of the state-of-the-art optical, ultrasound, and computational imaging approaches, has the potential to significantly impact imaging deep inside complex samples.

\section{Funding}
%%%% will generate automtically at prism platform for Optica submission: %%%
European Research Council (ERC) Horizon 2020 research and innovation program (grant no. 677909), Azrieli foundation, Israel Science Foundation (1361/18),Israeli Ministry of Science and Technology, National Science Foundation (1813848).

\section{Acknowledgments}
We thank Prof. Hagai Eisensberg for the q-switched laser and thank the Nanocenter at the Hebrew University, with special thanks to Dr. Itzik Shweky and Galina Chechelinsky, for fabrication the target samples.

%need to add: thank to NIST code for creating the targets.
%[1] The Nanolithography Toolbox, K. C. Balram, D. A. Westly, M. Davanco, K. E. Grutter, Q. Li, T. Michels, C. H. Ray, L. Yu,  R. J. Kasica,  C. B. Wallin, I. J. Gilbert, B. A. Bryce, G. Simelgor, J. Topolancik, N. Lobontiu , Y. Liu, P. Neuzil, V. Svatos, K. A. Dill, N. A. Bertrand, M. G. Metzler, G. Lopez, D. A. Czaplewski, L. Ocola, K. A. Srinivasan, S. M. Stavis, V. A. Aksyuk, J. A. Liddle, S. Krylov and B. R. Ilic, J. Res. Natl. Inst. Stand. 121, pp. 464-475 (2016).

\section{Disclosures}

\noindent\textbf{Disclosures.} The authors declare no conflicts of interest.

% Bibliography
\bibliography{reference}

\begin{thebibliography}{10}
\newcommand{\enquote}[1]{``#1''}

\bibitem{ntziachristos2010going}
V.~Ntziachristos, \enquote{Going deeper than microscopy: the optical imaging
  frontier in biology,} {\protect\JournalTitle{Nature methods}} \textbf{7},
  603--614 (2010).

\bibitem{pawley2006handbook}
J.~Pawley, \emph{Handbook of biological confocal microscopy}, vol. 236
  (Springer Science \& Business Media, 2006).

\bibitem{webb1996confocal}
R.~H. Webb, \enquote{Confocal optical microscopy,}
  {\protect\JournalTitle{Reports on Progress in Physics}} \textbf{59}, 427
  (1996).

\bibitem{huang1991optical}
D.~Huang, E.~A. Swanson, C.~P. Lin, J.~S. Schuman, W.~G. Stinson, W.~Chang,
  M.~R. Hee, T.~Flotte, K.~Gregory, C.~A. Puliafito \emph{et~al.},
  \enquote{Optical coherence tomography,} {\protect\JournalTitle{science}}
  \textbf{254}, 1178--1181 (1991).

\bibitem{fercher2003optical}
A.~F. Fercher, W.~Drexler, C.~K. Hitzenberger, and T.~Lasser, \enquote{Optical
  coherence tomography-principles and applications,}
  {\protect\JournalTitle{Reports on progress in physics}} \textbf{66}, 239
  (2003).

\bibitem{drexler2008optical}
W.~Drexler and J.~G. Fujimoto, \emph{Optical coherence tomography: technology
  and applications} (Springer Science \& Business Media, 2008).

\bibitem{jang2020deep}
M.~Jang, H.~Ko, J.~H. Hong, W.~K. Lee, J.-S. Lee, and W.~Choi, \enquote{Deep
  tissue space-gated microscopy via acousto-optic interaction,}
  {\protect\JournalTitle{Nature communications}} \textbf{11}, 1--11 (2020).

\bibitem{elson2011ultrasound}
D.~S. Elson, R.~Li, C.~Dunsby, R.~Eckersley, and M.-X. Tang,
  \enquote{Ultrasound-mediated optical tomography: a review of current
  methods,} {\protect\JournalTitle{Interface Focus}} \textbf{1}, 632--648
  (2011).

\bibitem{wang2004ultrasound}
L.~V. Wang, \enquote{Ultrasound-mediated biophotonic imaging: a review of
  acousto-optical tomography and photo-acoustic tomography,}
  {\protect\JournalTitle{Disease markers}} \textbf{19}, 123--138 (2004).

\bibitem{resink2012state}
S.~G. Resink, W.~Steenbergen, and A.~C. Boccara, \enquote{State-of-the art of
  acoust-optic sensing and imaging of turbid media,}
  {\protect\JournalTitle{Journal of biomedical optics}} \textbf{17}, 040901
  (2012).

\bibitem{wang2012photoacoustic}
L.~V. Wang and S.~Hu, \enquote{Photoacoustic tomography: in vivo imaging from
  organelles to organs,} {\protect\JournalTitle{science}} \textbf{335},
  1458--1462 (2012).

\bibitem{liu2015optical}
Y.~Liu, P.~Lai, C.~Ma, X.~Xu, A.~A. Grabar, and L.~V. Wang, \enquote{Optical
  focusing deep inside dynamic scattering media with near-infrared
  time-reversed ultrasonically encoded (true) light,}
  {\protect\JournalTitle{Nature communications}} \textbf{6}, 1--9 (2015).

\bibitem{wang2012deep}
Y.~M. Wang, B.~Judkewitz, C.~A. DiMarzio, and C.~Yang, \enquote{Deep-tissue
  focal fluorescence imaging with digitally time-reversed ultrasound-encoded
  light,} {\protect\JournalTitle{Nature communications}} \textbf{3}, 1--8
  (2012).

\bibitem{si2012fluorescence}
K.~Si, R.~Fiolka, and M.~Cui, \enquote{Fluorescence imaging beyond the
  ballistic regime by ultrasound-pulse-guided digital phase conjugation,}
  {\protect\JournalTitle{Nature photonics}} \textbf{6}, 657--661 (2012).

\bibitem{si2012breaking}
K.~Si, R.~Fiolka, and M.~Cui, \enquote{Breaking the spatial resolution barrier
  via iterative sound-light interaction in deep tissue microscopy,}
  {\protect\JournalTitle{Scientific reports}} \textbf{2}, 748 (2012).

\bibitem{ruan2014iterative}
H.~Ruan, M.~Jang, B.~Judkewitz, and C.~Yang, \enquote{Iterative time-reversed
  ultrasonically encoded light focusing in backscattering mode,}
  {\protect\JournalTitle{Scientific reports}} \textbf{4}, 7156 (2014).

\bibitem{judkewitz2013speckle}
B.~Judkewitz, Y.~M. Wang, R.~Horstmeyer, A.~Mathy, and C.~Yang,
  \enquote{Speckle-scale focusing in the diffusive regime with time reversal of
  variance-encoded light (trove),} {\protect\JournalTitle{Nature photonics}}
  \textbf{7}, 300--305 (2013).

\bibitem{katz2019controlling}
O.~Katz, F.~Ramaz, S.~Gigan, and M.~Fink, \enquote{Controlling light in complex
  media beyond the acoustic diffraction-limit using the acousto-optic
  transmission matrix,} {\protect\JournalTitle{Nature communications}}
  \textbf{10}, 1--10 (2019).

\bibitem{cui2010implementation}
M.~Cui and C.~Yang, \enquote{Implementation of a digital optical phase
  conjugation system and its application to study the robustness of turbidity
  suppression by phase conjugation,} {\protect\JournalTitle{Optics express}}
  \textbf{18}, 3444--3455 (2010).

\bibitem{doktofsky2020acousto}
D.~Doktofsky, M.~Rosenfeld, and O.~Katz, \enquote{Acousto optic imaging beyond
  the acoustic diffraction limit using speckle decorrelation,}
  {\protect\JournalTitle{Communications Physics}} \textbf{3}, 1--8 (2020).

\bibitem{ruan2020fluorescence}
H.~Ruan, Y.~Liu, J.~Xu, Y.~Huang, and C.~Yang, \enquote{Fluorescence imaging
  through dynamic scattering media with speckle-encoded ultrasound-modulated
  light correlation,} {\protect\JournalTitle{Nature Photonics}} pp. 1--6
  (2020).

\bibitem{feng1988correlations}
S.~Feng, C.~Kane, P.~A. Lee, and A.~D. Stone, \enquote{Correlations and
  fluctuations of coherent wave transmission through disordered media,}
  {\protect\JournalTitle{Physical review letters}} \textbf{61}, 834 (1988).

\bibitem{bertolotti2012non}
J.~Bertolotti, E.~G. Van~Putten, C.~Blum, A.~Lagendijk, W.~L. Vos, and A.~P.
  Mosk, \enquote{Non-invasive imaging through opaque scattering layers,}
  {\protect\JournalTitle{Nature}} \textbf{491}, 232--234 (2012).

\bibitem{katz2014non}
O.~Katz, P.~Heidmann, M.~Fink, and S.~Gigan, \enquote{Non-invasive single-shot
  imaging through scattering layers and around corners via speckle
  correlations,} {\protect\JournalTitle{Nature photonics}} \textbf{8}, 784--790
  (2014).

\bibitem{judkewitz2015translation}
B.~Judkewitz, R.~Horstmeyer, I.~M. Vellekoop, I.~N. Papadopoulos, and C.~Yang,
  \enquote{Translation correlations in anisotropically scattering media,}
  {\protect\JournalTitle{Nature physics}} \textbf{11}, 684--689 (2015).

\bibitem{osnabrugge2017generalized}
G.~Osnabrugge, R.~Horstmeyer, I.~N. Papadopoulos, B.~Judkewitz, and I.~M.
  Vellekoop, \enquote{Generalized optical memory effect,}
  {\protect\JournalTitle{Optica}} \textbf{4}, 886--892 (2017).

\bibitem{maiden2009improved}
A.~M. Maiden and J.~M. Rodenburg, \enquote{An improved ptychographical phase
  retrieval algorithm for diffractive imaging,}
  {\protect\JournalTitle{Ultramicroscopy}} \textbf{109}, 1256--1262 (2009).

\bibitem{maiden2017further}
A.~Maiden, D.~Johnson, and P.~Li, \enquote{Further improvements to the
  ptychographical iterative engine,} {\protect\JournalTitle{Optica}}
  \textbf{4}, 736--745 (2017).

\bibitem{pham2019semi}
M.~Pham, A.~Rana, J.~Miao, and S.~Osher, \enquote{Semi-implicit relaxed
  douglas-rachford algorithm (sdr) for ptychography,}
  {\protect\JournalTitle{Optics express}} \textbf{27}, 31246--31260 (2019).

\bibitem{gross2005heterodyne}
M.~Gross, P.~Goy, B.~Forget, M.~Atlan, F.~Ramaz, A.~Boccara, and A.~Dunn,
  \enquote{Heterodyne detection of multiply scattered monochromatic light with
  a multipixel detector,} {\protect\JournalTitle{Optics letters}} \textbf{30},
  1357--1359 (2005).

\bibitem{edrei2016optical}
E.~Edrei and G.~Scarcelli, \enquote{Optical imaging through dynamic turbid
  media using the fourier-domain shower-curtain effect,}
  {\protect\JournalTitle{Optica}} \textbf{3}, 71--74 (2016).

\bibitem{freund1990looking}
I.~Freund, \enquote{Looking through walls and around corners,}
  {\protect\JournalTitle{Physica A: Statistical Mechanics and its
  Applications}} \textbf{168}, 49--65 (1990).

\bibitem{gardner2019ptychographic}
D.~F. Gardner, S.~Divitt, and A.~T. Watnik, \enquote{Ptychographic imaging of
  incoherently illuminated extended objects using speckle correlations,}
  {\protect\JournalTitle{Applied optics}} \textbf{58}, 3564--3569 (2019).

\bibitem{li2019imaging}
Z.~Li, D.~Wen, Z.~Song, T.~Jiang, W.~Zhang, G.~Liu, and X.~Wei,
  \enquote{Imaging correlography using ptychography,}
  {\protect\JournalTitle{Applied Sciences}} \textbf{9}, 4377 (2019).

\bibitem{peterson2016probe}
I.~Peterson, R.~Harder, and I.~Robinson, \enquote{Probe-diverse ptychography,}
  {\protect\JournalTitle{Ultramicroscopy}} \textbf{171}, 77--81 (2016).

\bibitem{idell1987image}
P.~S. Idell, J.~R. Fienup, and R.~S. Goodman, \enquote{Image synthesis from
  nonimaged laser-speckle patterns,} {\protect\JournalTitle{Optics letters}}
  \textbf{12}, 858--860 (1987).

\bibitem{fienup1978reconstruction}
J.~R. Fienup, \enquote{Reconstruction of an object from the modulus of its
  fourier transform,} {\protect\JournalTitle{Optics letters}} \textbf{3},
  27--29 (1978).

\bibitem{rodenburg2004phase}
J.~M. Rodenburg and H.~M. Faulkner, \enquote{A phase retrieval algorithm for
  shifting illumination,} {\protect\JournalTitle{Applied physics letters}}
  \textbf{85}, 4795--4797 (2004).

\bibitem{li2019image}
G.~Li, W.~Yang, H.~Wang, and G.~Situ, \enquote{Image transmission through
  scattering media using ptychographic iterative engine,}
  {\protect\JournalTitle{Applied Sciences}} \textbf{9}, 849 (2019).

\bibitem{zhou2020retrieval}
M.~Zhou, R.~Li, T.~Peng, A.~Pan, J.~Min, C.~Bai, D.~Dan, and B.~Yao,
  \enquote{Retrieval of non-sparse objects through scattering media beyond the
  memory effect,} {\protect\JournalTitle{Journal of Optics}} \textbf{22},
  085606 (2020).

\bibitem{divitt2020imaging}
S.~Divitt, D.~F. Gardner, and A.~T. Watnik, \enquote{Imaging around corners in
  the mid-infrared using speckle correlations,} {\protect\JournalTitle{Optics
  Express}} \textbf{28}, 11051--11064 (2020).

\bibitem{selb2002nonlinear}
J.~Selb, L.~Pottier, and A.~C. Boccara, \enquote{Nonlinear effects in
  acousto-optic imaging,} {\protect\JournalTitle{Optics letters}} \textbf{27},
  918--920 (2002).

\bibitem{freund1988memory}
I.~Freund, M.~Rosenbluh, and S.~Feng, \enquote{Memory effects in propagation of
  optical waves through disordered media,} {\protect\JournalTitle{Physical
  review letters}} \textbf{61}, 2328 (1988).

\bibitem{schott2015characterization}
S.~Schott, J.~Bertolotti, J.-F. L{\'e}ger, L.~Bourdieu, and S.~Gigan,
  \enquote{Characterization of the angular memory effect of scattered light in
  biological tissues,} {\protect\JournalTitle{Optics express}} \textbf{23},
  13505--13516 (2015).

\bibitem{liu2017focusing}
Y.~Liu, C.~Ma, Y.~Shen, J.~Shi, and L.~V. Wang, \enquote{Focusing light inside
  dynamic scattering media with millisecond digital optical phase conjugation,}
  {\protect\JournalTitle{Optica}} \textbf{4}, 280--288 (2017).

\bibitem{liu2016lock}
Y.~Liu, Y.~Shen, C.~Ma, J.~Shi, and L.~V. Wang, \enquote{Lock-in camera based
  heterodyne holography for ultrasound-modulated optical tomography inside
  dynamic scattering media,} {\protect\JournalTitle{Applied physics letters}}
  \textbf{108}, 231106 (2016).

\bibitem{shevlin2018phase}
F.~Shevlin, \enquote{Phase randomization for spatiotemporal averaging of
  unwanted interference effects arising from coherence,}
  {\protect\JournalTitle{Applied optics}} \textbf{57}, E6--E10 (2018).

\bibitem{laudereau2016ultrafast}
J.-B. Laudereau, A.~A. Grabar, M.~Tanter, J.-L. Gennisson, and F.~Ramaz,
  \enquote{Ultrafast acousto-optic imaging with ultrasonic plane waves,}
  {\protect\JournalTitle{Optics Express}} \textbf{24}, 3774--3789 (2016).

\bibitem{metzler2020deep}
C.~A. Metzler, F.~Heide, P.~Rangarajan, M.~M. Balaji, A.~Viswanath,
  A.~Veeraraghavan, and R.~G. Baraniuk, \enquote{Deep-inverse correlography:
  towards real-time high-resolution non-line-of-sight imaging,}
  {\protect\JournalTitle{Optica}} \textbf{7}, 63--71 (2020).

\end{thebibliography}



\section*{Supplementary material}

## Acousto-optic Ptychography: Supplemental document

### 1. EXPERIMENTAL SETUP

The full experimental setup is shown schematically in Supplementary Figure S1. It is a pulsed acousto-optic imaging (AOI) experiment, employing a camera-based digital phase-shifting holographic detection [1, 2]. The light source is a single longitudinal-mode 532$nm$ passive Q-switched laser (Standa STA-01-SH-4) providing < 1$ns$ duration pulses at $f_{rep} = 25kHz$ repetition rate. The laser output is split to a reference and object arm by a polarization beam splitter (PBS), with a half-wave plate (HWP) controlling the splitting ratio.

At the object arm the beam is expanded to a width of $\sim 4mm$, passed through an $f = 200mm$ focal length lens and a diffuser (two stacked $1^o$ holographic diffusers Newport) mounted on a computer-controlled rotating stage, generating a controlled speckle decorrelation-time in the illumination beam. The speckled beam was introduced into the water tank containing the target, placed between two scattering layers, in a sandwich configuration. The first scattering layer was a $5^o$ diffuser placed $\sim 1cm$ before the target, and the second layer was composed of two $5^o$ diffusers stacked together. The distance between the target to the second layer was $L = 5cm$. The targets were fabricated on 1.5mm thick glass slides coated with Ti and Ag. The Ti layer was 20$nm$ thick placed above an Ag coating of 100$nm$ thickness, fabricated using E-Beam Lithography.

An sCMOS camera (Andor Zyla 4.2) was placed at a distance of 9$cm$ from the second diffuser. For acousto-optic tagging, an ultrasound transducer with a central frequency of 25MHz (Olympus

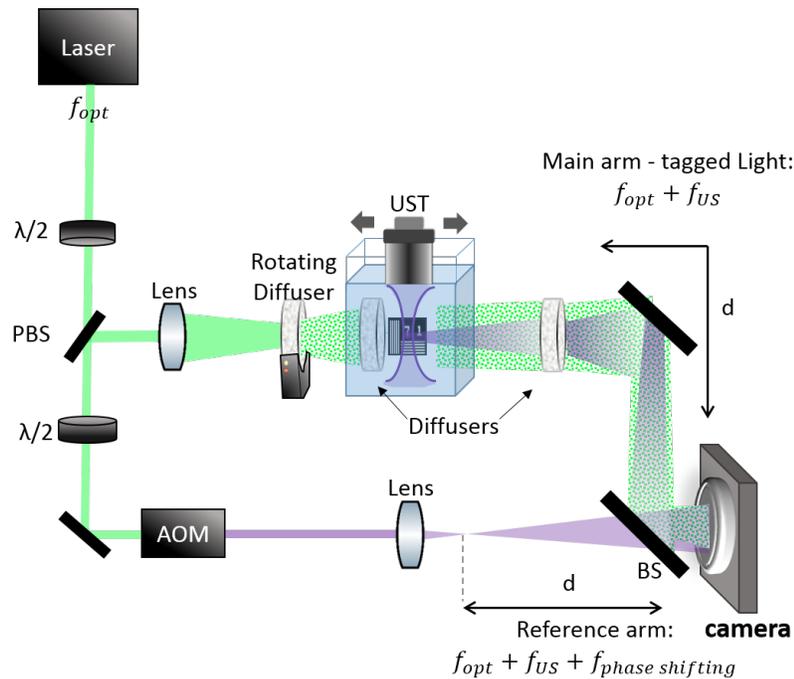

**Fig. S1.** The experimental setup. PBS - polarizing beam splitter. AOM - two acousto optic modulators. BD - beam dump. FG - function generator. AMP - amplifier. UST - ultrasonic transducer. $f_{opt}$ - laser optical frequency ($f_{opt}=c/\lambda$). $f_{US}$ - ultrasound driving frequency: $f_{US} = 25MHz + f_{rep}/2 + 1/T_{exposure}$, where $f_{rep}$ is the laser pulse repetition-rate, and $T_{exposure}$ is the camera exposure time

v324, focal length = 12.7mm, F# = 2) was used. The transducer was driven by 100ns-long sinusoidal pulses at a central frequency of $f_{us} = 25MHz + f_{rep}/2 + 1/T_{exposure}$, where $T_{exposure}$ is the camera exposure time. The pulses were generated by a function generator (Keysight 33600A) with a peak-to-peak amplitude of $2.4V_{pp}$, amplified by 40 dB by an RF power amplifier (Amplifier Research 25A250A), resulting in $240V_{pp}$ driving amplitude.

The ultrasound transducer was mounted on a computer-controlled motorized translation stage (Thorlabs), which allowed transverse horizontal scanning of the acoustic focus. The vertical position of the ultrasound focus was controlled by varying the relative delay between the acoustic pulses and the laser trigger pulses. To ensure accurate temporal synchronization of the $< 1ns$ laser pulses and to compensate for slow temporal drifts of the passively q-switched laser pulses timing from the laser trigger, a photodiode was used to trigger the acoustic pulse using the previous laser pulse.

At the reference arm, two acousto-optic modulators (AA Opto-Electronic MT80-A1) connected to RF amplifiers (Mini-Circuits ZHL-3a) were used to frequency shift the reference beam by $f_{AOM} = f_{US} + f_{\text{phase shifting}}/4$, where $f_{\text{phase shifting}} = f_{cam}/4$, and $f_{cam}$ is the sCMOS camera frame-rate.

The ultrasound frequency shift was chosen to be $f_{us} = 25MHz + f_{rep}/2 + 1/T_{exposure}$, to both efficiently reject the coherent interference of the unmodulated speckle background, which originates from the use of the laser pulses shorter than the ultrasound period and pulse width [3, 4], and to average over the sinusoidal spatial modulation of the ultrasound pulse. Efficient background rejection of the unmodulated light is obtained by adding a frequency-shift of $f_{rep}/2 = 12.5KHz$ [3], and the additional $1/T_{exposure}$ exploits the travelling ultrasound pulse to average the sinusoidal spatial modulation over a single ultrasound wavelength [2], effectively providing smooth ultrasound probe in the pulse propagation (vertical) direction (Fig. S2(a,c)). The additional frequency-shift of $f_{\text{phase shifting}} = 5Hz$ allows 4-frame phase-shifting holographic detection.

## 2. ULTRASOUND PROBE CHARACTERIZATION

A direct optical characterization of the ultrasound focus, as measured by removing the diffusers before the camera and placing an imaging lens, averaged over 200 different speckle illuminations is presented in Fig. S2. One-dimensional profiles of the ultrasound focus along the horizontal (x-axis) and vertical (y-axis) dimensions are presented in Fig. S2(b-c). Their measured full width at half max (FWHM) are $D_X = 149\mu m$, and $D_Y = 140\mu m$ (FWHM), respectively.

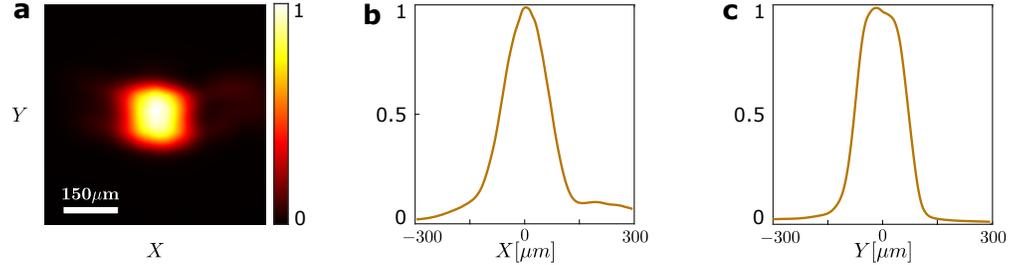

**Fig. S2.** Ultrasound probe characterization: (a) Direct image of the acoustic focus used in the experiments (Fig. 2-3 in the main text), averaged over 200 speckle realizations. (b) One-dimensional profile of the ultrasound focus along the lateral (horizontal) direction, obtained by summing over the columns of (a). (c) One-dimensional vertical profile of the ultrasound focus, obtained by summing over the rows of (a). scale-bar $150\mu m$.

## 3. EXPERIMENTAL CHARACTERIZATION OF THE SCATTERING LAYERS MEMORY-EFFECT RANGE

The memory effect range of the scattering layer through which imaging was performed in our experiments (two $5^o$ diffusers, Newport) was characterized by comparing the autocorrelation of the direct image of the extended object (Supplementary Fig. S3(a)) used in Fig. 2 in the main text, obtained using an imaging-lens replacing the scattering layer (Supplementary Figure S3(b)) to the autocorrelation of the object obtained using correlography through the scattering layer but without acousto-optic tagging, following Edrei et al [5] (Supplementary Fig. S3(c)). The

memory-effect correlation at each shift at the object plane was estimated by calculating the ratio between the two autocorrelations, and taking the envelope of the resulting ratio.

Supplementary Fig. S3(d) presents the horizontal cross-section of the memory effect range, along with the acoustic probe cross-section and auto-correlation. The resulting memory-effect FWHM is $\sim 280\mu m$ at the target distance of $\sim 5cm$, i.e. an angular memory-effect range of $\sim 5.6 mrad = 0.32^o$. The requirement for acousto-optic ptychographic imaging (AOPI) is that the acoustic probe will be contained withing the memory effect range. In this way, each position can be retrieved using speckle correlation method, while the full object can extend beyond the memory effect range indefinitely. As can be seen from Fig. S3(d), in our experimental conditions while the extended object (Fig. 2) indeed extends beyond the memory effect, the ultrasound probe itself and its spatial autocorrelation are smaller than the memory-effect correlation range, as required. In cases where the probe dimensions are similar to the memory effect range, one can digitally compensate the memory effect correlation reduction by normalizing the calculated autocorrelation through the scattering medium by the memory effect correlation function at each angle. Using this method will enable to relax the requirement on the probe size concerning the memory effect range, which might be significant when imaging in biological samples.

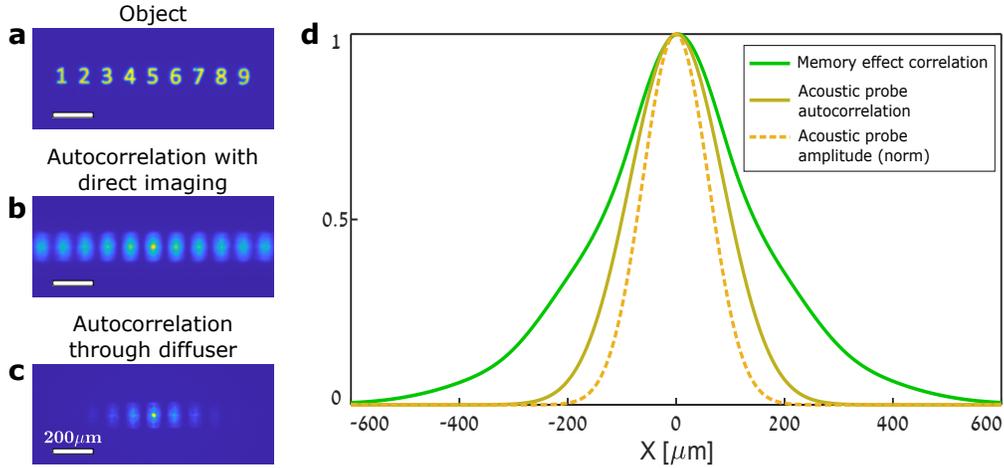

**Fig. S3.** Experimental characterization of the memory-effect range of the scattering layer used in our experiments by calculating the ratio between the autocorrelation of an extended object retrieved through the scattering layer to the autocorrelation of the object with direct imaging without the scattering layer. (a) The target object. (b) The object autocorrelation obtained with direct imaging without the scattering layer. (c) The object autocorrelation obtained through the scattering layer via correlography. (d) Horizontal cross-sections of: the experimentally measured memory-effect correlation as a function of translation at the object plane (bright green curve), the acoustic probe autocorrelation (dark green curve), and the acoustic probe amplitude (dashed curve). The memory effect correlation FWHM is $280\mu m$, , which is $\sim 3.5$ times smaller than the object size, but $\sim 2$ times larger than the ultrasound focus dimensions. Scale bars: $200\mu m$.

## 4. DATA PROCESSING FOR CALCULATING THE ULTRASOUND-GATED AUTOCORRELATIONS

Here we provide the technical details of our experimental autocorrelation calculation. As discussed in the main text, the complex field autocorrelation of the target can be calculated from a single camera frame obtained for a single speckle illumination [6]. However, this autocorrelation will be the autocorrelation of a speckled target, and thus will contain a sharp diffraction-limited peak representing the diffraction-limited speckle grain size. To overcome this, multiple camera frames captured under different speckle illuminations can be used to estimate the target incoherent intensity autocorrelation which, in the case of sufficient speckle averaging, is free from speckle artifacts via correlography[5–8]. The use of multiple speckle realizations is advantageous also for improving the SNR of the autocorrelation estimation via ensemble averaging [6]. While for an infinite number of speckle patterns the diffraction-limited speckle autocorrelation peak will decrease to zero, when a finite amount of realizations are used, the autocorrelation will still

3

contain speckle features (see next section) and contain a sharp diffraction limited peak, which can be significantly larger than the object autocorrelation features. In our experiments, 150 different speckle realizations where used at each scan position, and the speckle-grain autocorrelation peak was reduced by replacing the peak intensity with an intensity that is $\sim$ 3-times larger than its neighbouring pixels, following [5].

For background noise removal and filtering, the peak-reduced autocorrelation was additionally filtered by a Tukey window with a width adjusted to the ultrasound focus autocorrelation width. Finally, the windowed autocorrelation was filtered by thresholding its Fourier magnitude, resulting in noise removal and spatial smoothing.

## 5. EFFECT OF THE NUMBER OF SPECKLE REALIZATIONS ON AUTO-CORRELATION ESTIMATION

To determine the number of required speckle realizations (illuminations) we have studied the effect of the number of speckle realizations on the retrieved autocorrelation. The results of this study are given in Fig. S4. As expected, a lower number of speckle realizations (Fig. S4.(d-f)) results in speckle artefacts in the autocorrelation estimate, harming the reconstruction fidelity.

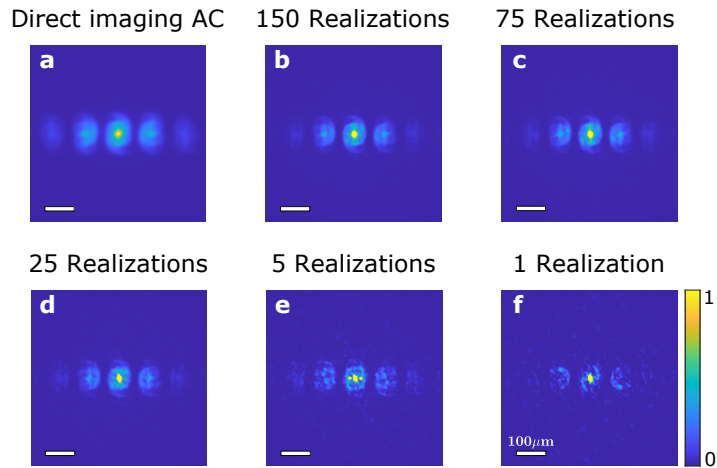

**Fig. S4.** Experimental investigation of autocorrelation estimation as a function of the number of speckle realizations (illuminations) .(a) The true autocorrelation of the extended target of Fig. 2 in the main text, tagged by the acoustic probe at a single position, as captured via direct imaging, averaged over 150 speckle realizations. (b) The correlography-retrieved autocorrelation of the target through the scattering layer, obtained from $N = 150$ different speckled fields. (c-f) Same as b, but using $N = 75, 25, 5, 1$ speckle realizations, respectively. Scale-bars $100 \mu m$.

## 6. CHARACTERIZATION OF THE DIFFUSER USED IN THE EXPERIMENT

When imaging through a scattering medium, it can be suspected that some of the imaging information is carried by the ballistic components that pass through the scattering medium along with the diffused components. To verify that in our experiments there is no significant ballistic component that allows imaging, and to study the scattering of the used diffusers, we performed an experiment that directly measures the scattering function of the scattering layer used in our experiments. To this end, we focused a beam from a collimated laser diode @520$nm$ (Thorlabs CPS520) on an sCMOS camera (Andor Zyla), using an f=100mm lens, and measured the focused intensity without and with the scattering layer placed in the path between the lens and the camera. The results of this experiments are shown in Fig. S5(a), and prove that the transmission through the scattering layer does not contain any significant ballistic component, as expected.

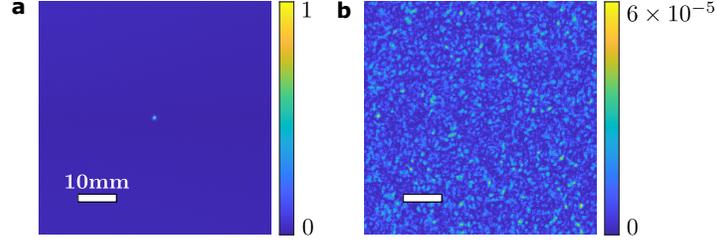

**Fig. S5.** Characterization of the scattering function of the scattering layer used in the experiment. (a) An image of a laser beam focused using an f=100mm lens without the scattering layer. (b) Same as a, but with the addition of the scattering layer after the focusing lens, showing no significant ballistic component case. Scale-bars: 10*mm*

## 7. PTYCHOGRAPHIC IMAGE RECONSTRUCTION PIPELINE

An iterative minimization algorithm was used to solve the joint-phase retrieval problem. We compared several algorithms from the ptychographic iterative engine (PIE) family [9, 10]. The best results on our experimental data were obtained using the regularized PIE (rPIE) algorithm [10], which was used for all reconstructions displayed in this work.

Below, we provide the details of the algorithmic steps for reconstructing the full objects from the set of estimated autocorrelations, which are calculated from the measured scattered patterns, as explained in the previous sections.

In a nutshell, the joint-phase retrieval algorithm follows a similar scheme to Fienup's Hybrid input output (HIO) algorithm for phase retrieval [11] but using a set of power spectra, which in our case are the Fourier magnitude of the measured autocorrelations, where the scanned acoustic probe acts as the isolated object constraint (spatial gate). The ptychographic algorithm starts by guessing an initial object, and estimation of the acoustic probe. During every iteration the algorithm enforces the object Fourier amplitude measurements and the estimated probe as constraints following the PIE update rules, until convergence. Since in our method we aim at reconstructing the target intensity pattern, we also enforced non-negativity constraint on the reconstructed object.

rPIE is a slightly modified advanced version of the original ePIE, differing in the update rules for the exit wave function, probe, and object [10]. For simplicity we present here the original ePIE algorithm. A full flowchart of our method, including both the autocorrelation estimation, and the ptychographic reconstruction, is given in Fig S6. At the m-th probe scan position and the j-th iteration of the algorithm we can write the exit wave function:

$$\psi_{j,m}(\mathbf{r}) \equiv O_j U_{j,m}(\mathbf{r} - r_m^{US}) \tag{S1}$$

Where $O_{j,m}(\mathbf{r})$ is the object and $U_{j,m}(\mathbf{r} - r_m^{US})$ is the probe illumination shifted by $r_m^{US}$ relative to **r** – a two dimensional coordinate along the ultrasound propagation and in perpendicular to it. The probe initial guess is estimated by a Gaussian profile in the horizontal axis and to a convolution between Gaussian and a rectangle profile in the vertical axis, from prior measurements of the acoustic probe (Supplementary section 2). In the next step, the algorithm constrains the Fourier amplitude of the exit wave function to the estimated Fourier amplitude from the object auto-correlation and keep the Fourier phase.

$$\psi'_{j,m}(\mathbf{r}) = \mathcal{F}^{-1}\{\frac{\mathcal{F}\{\psi_{j,m}(\mathbf{r})\}}{|\mathcal{F}\{\psi_{j,m}(\mathbf{r})\}|} \cdot |\mathcal{F}\{\psi_{0,m}(\mathbf{r})\}|\} \tag{S2}$$

Then, the algorithm use the new exit wave function to update first the object under the assumption that the probe is fixed and then the probe under the assumption that the object is fixed with the update terms:

$$O_{j+1} = O_j + \alpha \frac{U_j^*(\mathbf{r} - r_m^{US})}{|U_j(\mathbf{r} - r_m^{US})|_{max}^2} \cdot (\psi'_{j,m}(\mathbf{r}) - \psi_{j,m}(\mathbf{r})) \tag{S3}$$

$$U_{j+1} = U_j + \beta \frac{O_j^*(\mathbf{r} + r_m^{US})}{|O_j(\mathbf{r} + r_m^{US})|_{max}^2} \cdot (\psi'_{j,m}(\mathbf{r}) - \psi_{j,m}(\mathbf{r})) \tag{S4}$$

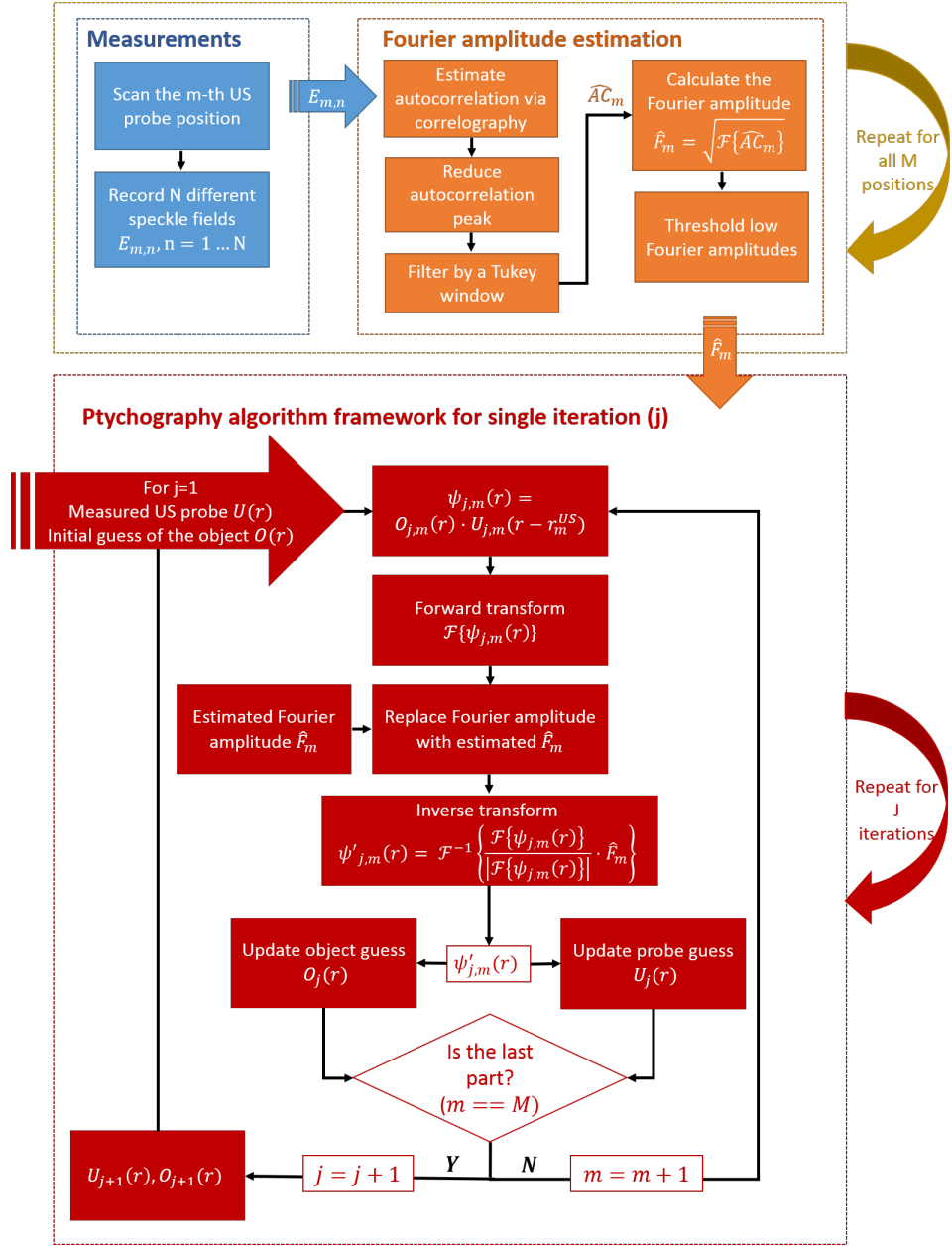

**Fig. S6.** Flow-chart of the acousto-optic ptychographic imaging (AOPI) reconstruction

Where $\alpha, \beta$ are weight parameters for the updated feedback. $\alpha$ is usually between $0.7 - 0.9$ and $\beta \approx 0.1$. When all M scan positions are gone through in a random order, the j-th ptychography iteration is completed.

## 8. NUMERICAL STUDY OF THE RECONSTRUCTION FIDELITY AS A FUNCTION OF NUMBER OF PROBE POSITIONS

The significant advantage of ptychography over individual reconstruction of the object at each probe position, is in the *joint* solution of the phase-retrieval problem, exploiting information contained in the overlap between neighboring probe positions [12]. To determine the required number of probe positions, or equivalently the overlap between neighbouring probe positions, we performed a numerical investigation of the influence of the number of probe positions on the reconstruction performance. The results of this study are shown in Figure S7.

For the numerical experiment a USAF-1951 resolution test chart with a minimal line width of 6.24$\mu m$ and an experimentally measured acoustic probe (Fig. S2) were used. The conventional AOI reconstruction from 868 probe positions is presented in Fig. S7(a) and shows a low-resolution reconstruction, without the ability to distinguish even the largest features of the target. An ideal AOPI reconstruction from 868 scans assuming infinite number of speckle averaging (i.e. effectively incoherent illumination) is presented in Fig. S7(b), where each autocorrelation was estimated from a single incoherent illumination, free from speckled pattern, and show a very-well resolved features with high-resolution reconstruction of the target. Fig. S7(c-f) show the AOPI reconstruction using 150 speckle realizations in each probe position, for a different number of probe positions: 868 scans (89% overlap between neighboring positions), 460 scans (80% overlap), 272 scans (67% overlap), and 72 scans (24% overlap), respectively.

To quantify the reconstruction fidelity, we calculate the structural similarity index (SSIM)[13] between each reconstruction and the target object. The graph in Fig. S7(g) plots the SSIM index as a function of number of probe positions. For the AOPI reconstruction using 150 speckle realizations, the reconstruction fidelity does not reduce until an overlap lower than 67% (i.e. 272 scans) is used. In our experiments an overlap of approximately 90% was used.

**REFERENCES**

1. M. Atlan, B. Forget, F. Ramaz, A. Boccara, and M. Gross, "Pulsed acousto-optic imaging in dynamic scattering media with heterodyne parallel speckle detection," Opt. letters **30**, 1360–1362 (2005).
2. D. Doktofsky, M. Rosenfeld, and O. Katz, "Acousto optic imaging beyond the acoustic diffraction limit using speckle decorrelation," Commun. Phys. **3**, 1–8 (2020).
3. K. Si, R. Fiolka, and M. Cui, "Fluorescence imaging beyond the ballistic regime by ultrasound-pulse-guided digital phase conjugation," Nat. photonics **6**, 657–661 (2012).
4. Y. M. Wang, B. Judkewitz, C. A. DiMarzio, and C. Yang, "Deep-tissue focal fluorescence imaging with digitally time-reversed ultrasound-encoded light," Nat. communications **3**, 1–8 (2012).
5. E. Edrei and G. Scarcelli, "Optical imaging through dynamic turbid media using the fourier-domain shower-curtain effect," Optica **3**, 71–74 (2016).
6. P. S. Idell, J. R. Fienup, and R. S. Goodman, "Image synthesis from nonimaged laser-speckle patterns," Opt. letters **12**, 858–860 (1987).
7. O. Salhov, G. Weinberg, and O. Katz, "Depth-resolved speckle-correlations imaging through scattering layers via coherence gating," Opt. letters **43**, 5528–5531 (2018).
8. Z. Li, D. Wen, Z. Song, T. Jiang, W. Zhang, G. Liu, and X. Wei, "Imaging correlography using ptychography," Appl. Sci. **9**, 4377 (2019).
9. A. M. Maiden and J. M. Rodenburg, "An improved ptychographical phase retrieval algorithm for diffractive imaging," Ultramicroscopy **109**, 1256–1262 (2009).
10. A. Maiden, D. Johnson, and P. Li, "Further improvements to the ptychographical iterative engine," Optica **4**, 736–745 (2017).
11. J. R. Fienup, "Reconstruction of an object from the modulus of its fourier transform," Opt. letters **3**, 27–29 (1978).
12. O. Bunk, M. Dierolf, S. Kynde, I. Johnson, O. Marti, and F. Pfeiffer, "Influence of the overlap parameter on the convergence of the ptychographical iterative engine," Ultramicroscopy **108**, 481–487 (2008).
13. Z. Wang, A. C. Bovik, H. R. Sheikh, and E. P. Simoncelli, "Image quality assessment: from error visibility to structural similarity," IEEE transactions on image processing **13**, 600–612 (2004).

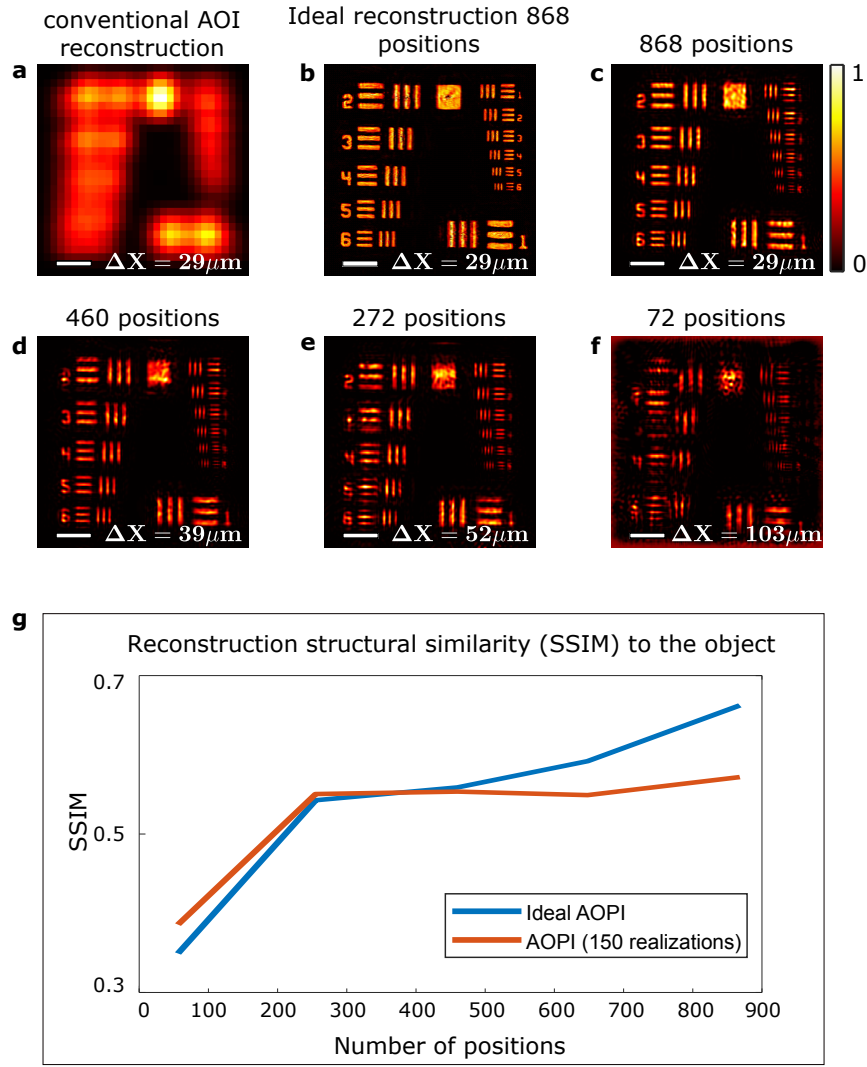

**Fig. S7.** Investigation of numerical AOPI reconstruction quality as a function of the number of probe positions (a varying overlap between neighbouring probes). (a) Conventional AOI reconstruction from 868 scans. (b) Ideal AOPI reconstruction from 868 scans (89% overlap between neighboring positions), using an infinite number of speckle realizations (incoherent illumination). (c-f) AOPI reconstructions using different number of probe positions: (c) 868 scans (89% overlap between neighboring positions, (d) 460 scans (80% overlap), (e) 272 scans (67% overlap), and (f) 72 scans (24% overlap). (g) quantification of reconstruction fidelity via the structural similarity index (SSIM) between each AOPI reconstruction and the target obejct. Scale-bars: $150 \mu m$.



\end{document}